\documentclass[12pt]{article}

\usepackage{newtxtext}
\usepackage{amssymb}
\usepackage{newtxmath}

\usepackage{graphicx}
\usepackage{siunitx}

\usepackage{braket}
\usepackage{multirow}

\usepackage{hyperref}

\usepackage[letterpaper,margin=1in]{geometry}

\renewenvironment{abstract}
	{\quotation}
	{\endquotation}

\date{}

\makeatletter
\renewcommand{\fnum@figure}{\textbf{Figure \thefigure}}
\renewcommand{\fnum@table}{\textbf{Table \thetable}}
\makeatother

\usepackage{scicite}

\usepackage{url}

\def\scititle{
	Multilayer Crystal Field states from locally broken centrosymmetry
}
\title{\bfseries \boldmath \scititle}

\author{
	Owen Moulding$^{1,2}$,
	Makoto Shimizu$^{3}$,
	Amit Pawbake$^{4}$, 
        Yingzheng Gao$^{1}$,\and
        Sitaram Ramakrishnan$^{1}$,
        Gaston Garbarino$^{5}$,
        Nubia Caroca-Canales$^{2}$,
        J\'er\^ome Debray$^{1}$,\and
        Cl\'{e}ment Faugeras$^{4}$,
        Christoph Geibel$^{2}$,
        Youichi Yanase$^{3}$,
        Marie-Aude M\'{e}asson$^{1\ast}$\and
	\small$^{1}$Institut N\'{e}el CNRS/UGA UPR2940, 25 Rue des Martyrs, 38042 Grenoble, France.\and
	\small$^{2}$Max-Planck Institute for Chemical Physics of Solids, 01187 Dresden, Germany.\and
    \small$^{3}$Department of Physics, Graduate School of Science, Kyoto University, Kyoto 606-8502, Japan.\and
    \small$^{4}$LNCMI, UPR 3228, CNRS, EMFL, Universit\'{e} Grenoble Alpes, 38000 Grenoble, France.\and
    \small$^{5}$European Synchrotron Radiation Facility, 71 Avenue des Martyrs, 38000 Grenoble, France.\and
	\small$^\ast$Corresponding author. Email: marie-aude.measson@neel.cnrs.fr\and
	}

\begin{document} 

\maketitle

\begin{abstract} \bfseries \boldmath
Local charge, spin, or orbital degrees of freedom with intersite interactions are oftentimes sufficient to construct most quantum orders. This is conventionally true for $f$-electron systems, where the extent of the $f$-electrons and their associated crystal-electric-field (CEF) states are strongly localized. Here, polarized Raman spectroscopy measurements of a locally non-centrosymmetric compound, CeCoSi, unveil more CEF excitations than expected in the local model. We interpret this as experimental evidence for the entanglement of CEF states between cerium layers. This composite sublattice, spin, and orbital degree of freedom provides an unconsidered means to form novel orders, not only in this system, but in any system exhibiting globally preserved yet locally broken centrosymmetry.
\end{abstract}

\noindent
\newpage
Degrees of freedom are the building blocks upon which ordered states are built in quantum materials. Common degrees of freedom include charge, spin, and orbital states, which when they interact with one another, give rise to quantum orders. However, these degrees of freedom are not necessarily sufficient to describe more exotic orderings, and multisite effects are largely overlooked, especially in $f$-electron systems due to their localized nature. The open question is, can we combine multisite physics and quantum ordering in conventionally localized materials? Globally centrosymmetric systems with broken local centrosymmetry intrinsically provide such a mechanism with a novel sublattice degree of freedom \cite{bauer2012_1}, which also induces an alternating Rashba asymmetric spin-orbit coupling (ASOC). In recent years, these compounds have been shown to beget unusual `hidden' Dresselhaus/Rashba phenomenon \cite{Dresselhaus1955_1,Rashba2015_1,Zhang2014_1}; most notably, parity-mixing superconductivity \cite{Fischer2011_1,Maruyama2012_1,Khim2021_1,Fischer2022_1, Nakamura2017}, and exotic multipolar states \cite{Schmidt2024_1,Hafner2022_1,Yatsushiro2022_1,Khim2021_1}.

CeCoSi belongs to this family of globally centrosymmetric structures with locally broken centrosymmetry \cite{Tanida2019_1}, and
hosts a hidden-order (HO) state below $T_0\approx$12\,K whose nature and degree of freedom are still puzzling \cite{Lengyel2013_1,Manago2021_1,Tanida2018_1,Tanida2019_1,Hidaka2022_1,Manago2023_1}. The origin of the HO state remains obscured, though what is clear is that within the HO state there is no conventional magnetism on the cerium atoms, as evidenced by inelastic neutrons scattering (INS) \cite{Nikitin2020_1}, nuclear-magnetic-resonance (NMR), and nuclear-quadrupolar-resonance (NQR) \cite{Manago2021_1, Manago2023_1}. Despite this, a magnetic interaction pervades the HO state, as $T_0$ is enhanced under a magnetic field \cite{Tanida2019_1, Hidaka2022_1}, and a magnetic field even modifies the low-temperature crystal structure of the HO state \cite{Matsumura2022_1}. In $f$-electrons systems, the fundamental orbital ingredient comes from the crystal-electric-field (CEF) levels, from which many higher-rank multipolar orders have been proposed for the origin of the HO state \cite{Yatsushiro2020_1,Yatsushiro2020_2, Yatsushiro2022_1, Yamada2024_1}. Vitally, such states can only form if there is a sufficient number of degrees of freedom provided by the CEF levels with compatible symmetries and meaningful energy scales. Here, we probe the CEF excitations of CeCoSi and provide evidence of novel entangled multisite states consisting of spin, orbital, and sublattice degrees of freedom.

\section*{Multilayer locally non-centrosymmetric structure}

\begin{figure}%
\centering
\includegraphics[width = 1\columnwidth]{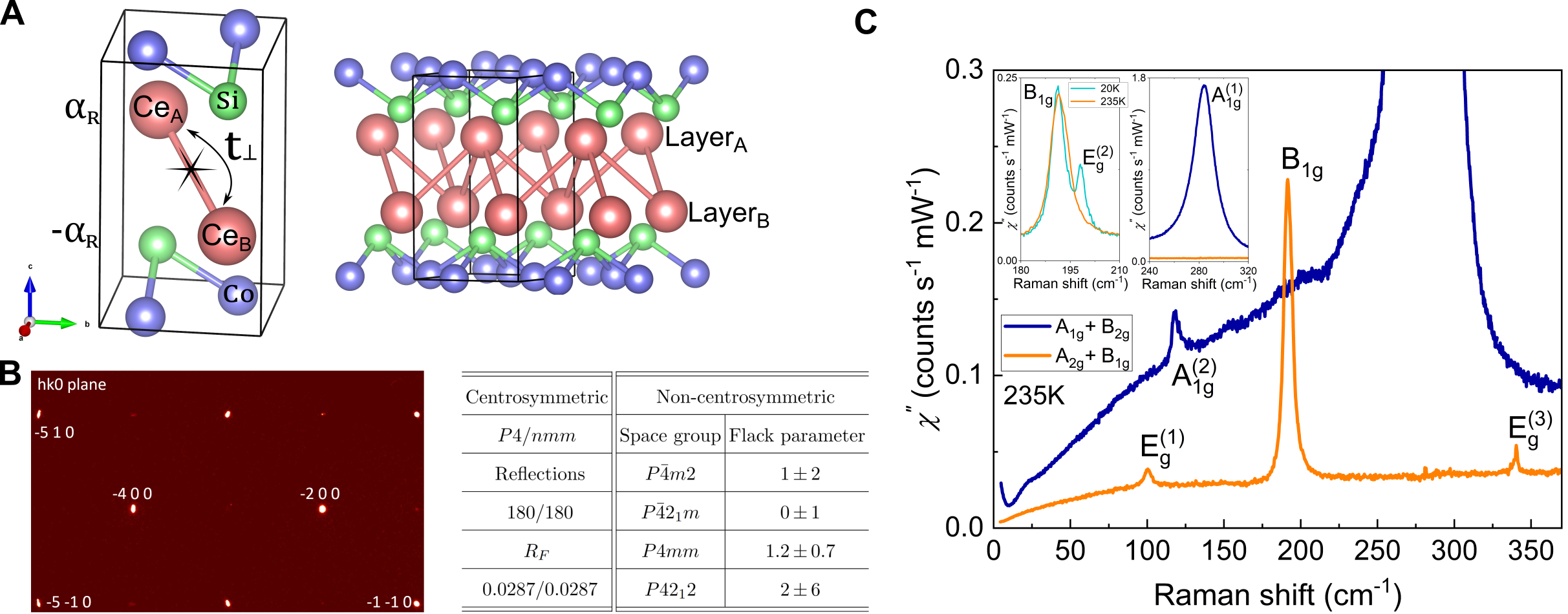}
\caption{\textbf{Locally non-centrosymmetric and multi-layer structure of  CeCoSi.} (\textbf{A}) Crystal structure of CeCoSi with a star indicating the local-inversion centre. $\alpha_R$ and $t_\perp$ indicate the alternating Rashba ASOC on each cerium layer, and the interlayer hopping term, respectively. Illustration of the two layer cerium structure of CeCoSi. (\textbf{B}) Reconstructed $hk$0 plane at room temperature from XRD measurements. Structural fitting is consistent with a $P$4/$nmm$ globally centrosymmetric structure with locally broken centrosymmetry at the cerium site. Table provides statistical information of the refined $P4/nmm$ structure, alongside the Flack parameters for the non-centrosymmetric subgroups. Values for reflections and $R_F$ correspond to observed/expected, respectively. (\textbf{C}) Raman spectra at high temperature in two symmetry configurations. Inset shows an additional low temperature spectrum highlighting the second E$_g$ mode. Labels show all the expected phonons of CeCoSi and their symmetry assignments.} %
\label{fig:Fig_1}%
\end{figure}

CeCoSi is known to contain two cerium ions per unit cell, which are related by global inversion symmetry, while the inversion centre is not situated on either of these ions. As a result, the compound is globally centrosymmetric with an absence of local centrosymmetry on the cerium ions. The cerium ions are separated by 3.8226\,\si{\angstrom} and form two distinct layers, as shown in Fig.\,\ref{fig:Fig_1}.A. We carefully checked the crystal structure of our sample at the ID15B beamline at the ESRF synchrotron, and we confirmed it to be the tetragonal $P$4/$nmm$ structure shown in Fig.\,\ref{fig:Fig_1}.A with lattice parameters $a$=4.0549(4)\,\si{\angstrom} and $c$=6.9841(7)\,\si{\angstrom} and atomic coordinates in the supplementary material (SM). Fig.\,\ref{fig:Fig_1}.B shows the reconstruction of the $hk$0 plane with strong reflections consistent with the $P$4/$nmm$ structure, and the statistical parameters determined for the refined $P4/nmm$ structure in the table. For the purpose of validating the crystal structure as globally centrosymmetric $P$4/$nmm$, we refined the structure with the non-centrosymmetric tetragonal subgroups of $P$4/$nmm$: $P\bar{4}m$2, $P\bar{4}2_{1}m$, $P$4$mm$, and $P$4$2_{1}$2. For each structure, the Flack parameter was determined, which accounts for inversion twinning. Physical Flack parameters should take values from 0 to 1 with these extremes implying pure untwinned structures. All the presented values and associated errors reported in the table are unphysical, which clearly supports the preservation of the centre of inversion. Further crystallographic information evidencing the refined $P4/nmm$ structure is shown in the SM. When all is taken into consideration, we conclude that the structure is unequivocally globally-centrosymmetric $P4/nmm$ in agreement with previous works \cite{Lengyel2013_1,Kawamura2020_1,Kawamura2022_1,Matsumura2022_1}.

From the known crystal group symmetry and the Wyckoff positions, we expect six Raman-active phonons with symmetries 2A$_{1g}\oplus$B$_{1g}\oplus$3E$_{g}$. Fig.\,\ref{fig:Fig_1}.C shows the spectra of two different symmetry configurations and all of the expected phonons are observed: the two A$_{1g}$ phonons and the one B$_{1g}$ are clearly observed, whilst the three E$_g$ phonons are weak but clear. Due to its proximity to the B$_{1g}$ phonon, the E$_g^{(2)}$ becomes resolvable at low temperatures as shown in the inset of Fig.\,\ref{fig:Fig_1}.C. The observance of the expected phonons is consistent with the structural analysis, and lays the foundation for the observance of the CEF excitations, which emerge at low temperature.

\section*{Crystal Field Modes and the failure of the standard localized model}

\begin{figure}%
\centering
\includegraphics[width = 1\columnwidth]{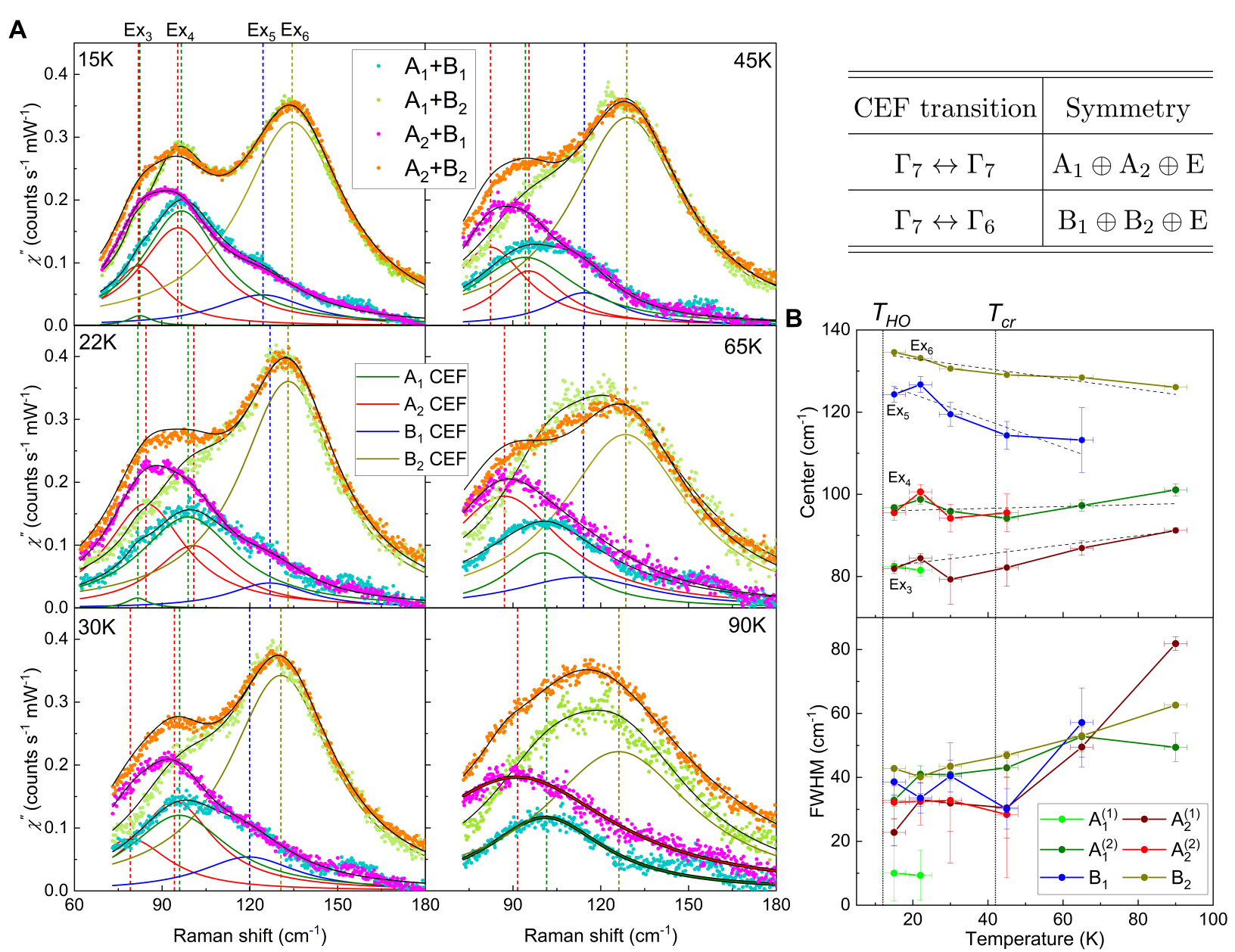}
\caption{\textbf{Raman signatures of the Crystal Electric Field modes in CeCoSi.} (\textbf{A}) Raman spectra of the CEF modes in four different symmetry decompositions at different temperatures after subtracting the phononic and Drude contributions. Solid lines show the individual symmetry responses of the CEF modes. Vertical dashed lines show the positions of the CEF modes. (\textbf{B}) Temperature dependence of the energy and full-width-half-maximum (FWHM) of the CEF modes. Dashed lines are guides to the eye. Dotted lines show the hidden-order transition ($T_{HO}$) and cross-over ($T_{cr}$) temperatures. The table show the symmetries of the expected CEF transitions.} %
\label{fig:Fig_2}%
\end{figure}%

In CeCoSi, like in many Ce-based strongly correlated compounds, the cerium electronic configuration is $4f^1$ (Ce$^{3+}$) \cite{Lengyel2013_1, Tanida2019_1}. In a tetragonal $C_{4v}$ environment, the six states of the cerium ions with total angular momentum $J$=5/2 are split into three Kramer's doublets with symmetries given by $2\Gamma_7 \oplus \Gamma_6$. In a purely localized model, there are then two excitations whose Raman activity is shown in the table of Fig.\,\ref{fig:Fig_2}. Using polarized Raman spectroscopy, we observe the CEF excitations below 100\,K. After subtracting the phonon and electronic background contributions (see SM), we obtained the temperature-dependent CEF spectra shown in Fig.\,\ref{fig:Fig_2}.A. We ascertained the number of CEF peaks and their symmetries by self-consistently fitting the four symmetry decompositions with the procedure outlined in the SM. The vertical dotted lines in Fig.\,\ref{fig:Fig_2}.A indicate the energies of the peaks, according to which we can sort these peaks into four groups corresponding to different CEF excitations. At 15\,K, the two first groups are at 82(2)\,\si{\centi \meter ^{-1}} and 96(2)\,\si{\centi \meter ^{-1}} and both composed of A$_{1}\oplus$A$_{2}$ symmetries. The third group consists of a single B$_{1}$ mode at 124(2)\,\si{\centi \meter ^{-1}}, while the final group is comprised of a single B$_{2}$ mode at 134.6(1)\,\si{\centi \meter ^{-1}}. These four groups of peaks remain energetically distinct up to 65\,K.

First and foremost, we observe four excitations as opposed to the expected two from a purely localized model. Concerning previous measurements, the energy range of these excitations is consistent with that determined by INS measurements \cite{Nikitin2020_1} and the Schottky anomaly observed in specific heat \cite{Tanida2019_1}. We shall return to the number of excitations later, but for now, we shall address the general CEF scheme, which is currently contentious. From our measurements, we observe a very clear symmetry dependence of the four excitations, which can be paired into A$_{1}$ and A$_{2}$ peaks at lower energy, and  B$_{1}$ and B$_{2}$ at higher energy. From the table in Fig.\,\ref{fig:Fig_2} and the energy ordering of the peaks, we infer that the CEF levels are ordered from lowest to highest energy as $\Gamma_7$, $\Gamma_7$, and $\Gamma_6$. Thus, we unambiguously determine the ordering of the CEF energy levels to be in agreement with Nikitin \textit{et al.} \cite{Nikitin2020_1}, and contrary to previous calculations that have predicted \cite{Yamada2024_1} or used \cite{Yatsushiro2020_1,Yatsushiro2020_2, Yatsushiro2022_1} other energetic orderings of the CEF levels. 

Returning to the four excitations, we eliminate two potential origins, namely structural and electronic transitions. For the former, there is no evidence of any change in the lattice in the temperature range measured from previous measurements \cite{Kawamura2020_1, Nikitin2020_1, Matsumura2022_1}, nor in the phononic response of our Raman measurements. For the latter, four electronic transitions have been observed over the entire temperature range: the antiferromagnetic regime below 9\,K \cite{Lengyel2013_1,Tanida2019_1, Nikitin2020_1}, the hidden-order transition at 12\,K ($T_{HO}$) \cite{Tanida2019_1}, the Kondo temperature below 17\,K \cite{Lengyel2013_1}, and the crossover temperature at 42\,K ($T_{cr}$) \cite{Lengyel2013_1}. From Fig.\,\ref{fig:Fig_2}.B, the four excitations exist much above all of these temperatures and therefore are an innate property of the normal state. Historically, a difference was observed in the number of expected CEF excitations in a few Ce-based compounds, such as CeAl$_2$ \cite{Guntherodt1983_1,Loewenhaupt1979_1,Loewenhaupt2003_1}, CeCuAl$_3$ \cite{Adroja2012_1}, or CeAuAl$_3$ \cite{cermak2019_1}. The appearance of these additional modes, so-called vibronic, requires the existence of a phonon with a compatible symmetry to the CEF excitation and in the same energy range to allow for strong magnetoelastic coupling. However, in CeCoSi, only the E$_g$ phonon at 100\,\si{\centi \meter ^{-1}} could couple to the energetically close $\Gamma_7 \leftrightarrow \Gamma_7$ excitation, and not the  $\Gamma_7 \leftrightarrow \Gamma_6$. Thus, magnetoelastic coupling cannot be the cause of the four excitations (see further details in the SM).

\section*{Multilayer Crystal Electric Field and Quantum orders}

\begin{figure}%
\centering
\includegraphics[width = 1\columnwidth]{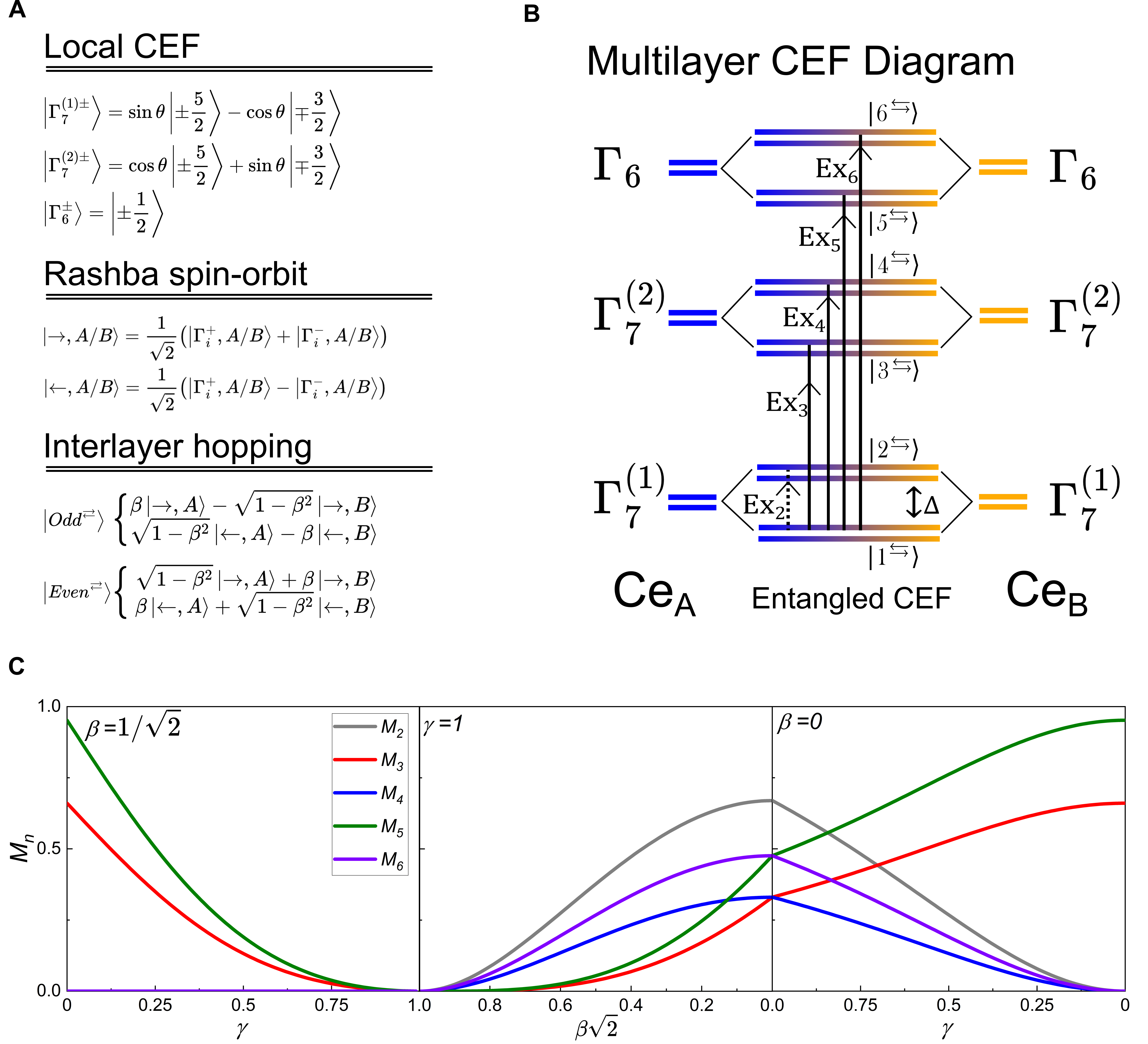}
\caption{\textbf{Multilayer Crystal Field model.} (\textbf{A}) Equations describing the purely localized CEF states of both cerium sites, the Rashba polarized states, and the final multilayer CEF states composed of entangled Rashba polarized states. $\beta$ is a function of $\alpha_R$ and $t_\perp$ (see main text). (\textbf{B}) Multilayer crystal field diagram of the two cerium layers.  Arrows correspond to the excitations (Ex$_i$) as measured. (\textbf{C}) Raman matrix element ($M_n$) from the ground state to the $n^{th}$ excited state at $T=$0\,K, encapsulating the conservation of the quantum number $m_j$ during the Raman processes, as functions of $\beta$ and $\gamma$. The latter describes the probability ratio of the inter- vs intra-layer transitions. Here $\theta=17.8 \si{\degree}$. As long as there are transitions between the two layers and the Rashba ASOC is present, all Raman transitions are allowed.}
\label{fig:Fig3}%
\end{figure}%

Here we show that the four excitations emerge from an extended CEF scheme based on the two cerium layers, namely a multilayer crystal field, which originates from the simultaneous presence of ASOC and interlayer coupling. For locally non-centrosymmetric layered systems, the ASOC sign ($\alpha_R$) alternates between the cerium layers (as shown in Fig.\,\ref{fig:Fig_1}.A), which leads to a layer-dependent spin polarization of the Fermi surface \cite{Maruyama2012_1,Zhang2014_1}. One consequence is that the original Kramer's doublets break, but new interlayer Kramer's doublets form \cite{Maruyama2012_1}. These new states are then protected by the remaining global inversion symmetry of the crystal and possess the spin and sublattice composite degree of freedom \cite{Maruyama2012_1}. With the addition of interlayer hopping ($t_\perp$), electrons with the same spin can hop between the different cerium layers. Overall, this results in new spin-polarized bonding/antibonding entangled states composed of interlayer Kramer's doublets. 

We extend this pseudospin model to a more realistic regime which possesses more complex CEF schemes. Starting with the purely localized $m_j$ states on the cerium $A$ and $B$ layers, which are stated in Fig.\,\ref{fig:Fig3}.A and identical by inversion symmetry, we define the single-site wavefunction ($\ket{\leftarrow, A/B}$ and $\ket{\rightarrow, A/B}$ for $A$ or $B$ layers), originating from the ASOC. Then, we construct entangled states between the two layers, as shown in Fig.\,\ref{fig:Fig3}.A. This admixture of states is parametrized by $\beta$, which is dependent on the ratio $\alpha_R/t_\perp$, the relative strength of the ASOC versus interlayer hopping, and given by,
\begin{equation}
   \beta =\frac{1}{\sqrt{1 + \left(  \frac{\alpha_R}{t_\perp}  + \sqrt{\left(\frac{\alpha_R}{t_\perp}\right)^2 + 1}  \right)^2}}.
   \label{eq:beta}
\end{equation}
The net result is that the three distinct CEF levels from each cerium ion (six in total but degenerate by inversion symmetry) hybridize to create six new CEF levels of bonding/antibonding states. Conceptually, this is similar to the formation of a molecule from localized orbitals, but here localized CEF levels form the entangled multilayer diagram in Fig.\,\ref{fig:Fig3}.B. This produces five Raman excitations from the ground state to the excited states of this multilayer CEF scheme. The lowest energy excitation (Ex$_2$) between the ground state $\ket{1^{\leftrightarrows}}$ and the first excited state $\ket{2^{\leftrightarrows}}$ across $\Delta$ is not currently observed. It may simply be too low in energy, i.e. below 5\,\si{\centi \meter ^{-1}} which was the lowest measurable energy in this work. Alternatively, thermal occupation would weaken and broaden the excitation to the point where we could not measure it. In the normal state at 15\,K, we evaluate a threshold for the observation of an excitation to be at least 10\,\si{\centi \meter ^{-1}} corresponding to 30\,\% occupation. The energy is of the same magnitude as the energy difference between Ex$_3$/Ex$_4$ and Ex$_5$/Ex$_6$. Indeed, we expect $\Delta$ to be comparable, but no symmetry constraint applies.  

To re-address the selection rules concerning our multilayer model, we observe that the selection rules for the symmetries of the excitations are the same as those between the localized CEF states, thus, the new multilayer states seem to inherit the symmetries of the original localized states. Next, we consider the selection rules of the Raman process constrained by the conservation of $m_j$ ($\Delta m_j=0, \pm 1$) \cite{Cardona2000}. We constructed a toy-model that describes the Raman scattering between the entangled CEF states of the multilayer diagram in Fig.\,\ref{fig:Fig3}.B. In addition to $\beta$, we include a parameter $\gamma=D/S$ with $D$ and $S$ weighting the inter- and intra-layer Raman transitions, respectively. For further details, see SM. When ASOC and interlayer transitions are both present, the five CEF excitations are activated, as shown in Fig.\,\ref{fig:Fig3}.C. The even-parity state transitions are only forbidden in the absence of ASOC ($\alpha_R$=0, $\beta=1/\sqrt{2}$, Fig.\,\ref{fig:Fig3}.C) alternatively, with forbidden interlayer transitions. These constraints are removed when we consider thermal occupation of the first excited state, see SM. All of these results are valid regardless of the value of $\theta$, which describes the composition of the localized $\Gamma_7$ states, except for two unlikely cases: perfectly balanced mixing of $\pm$5/2 and $\mp$3/2 $m_j$-states ($\theta=\pi/4$), or a pure $m_j$ state ($\theta=\pi/2$). Neither of these cases are pertinent to CeCoSi based on our DFT calculations (in SM), or INS measurements giving $\theta=17.8$\si{\degree} \cite{Nikitin2020_1}. To quantify the ratio of ASOC to interlayer hopping, we evaluate the Zeeman shift of the novel CEF states using high-magnetic field Raman measurements. We estimate $\frac{\alpha_R}{t_{\perp}}= 1.4$ ($\beta$=0.3(1)) (SM), and we postulate that CeCoSi being in this intermediate regime of balance between ASOC and interlayer hopping gives rise to the observed exotic phenomenon. 

Our proposed CEF scheme provides a hitherto unconsidered means to construct exotic quantum orderings. If we only consider the local degrees of freedom within the ground state Kramer's doublet of the multilayer CEF, we can form the standard charge and magnetic dipole orders. By including the quasi-quartet via thermal population of the first excited state across $\Delta$, five types of long-range order can arise: even/odd-parity electric and magnetic multipolar order, and magnetic toroidal order. These have also been called augmented or cluster multipoles \cite{Suzuki2018_1,Kusunose2020_1}. The even-parity electric and magnetic order preserve the global inversion symmetry, while the odd-parity electric dipole order and the magnetic monopole/toroidal order spontaneously break the inversion symmetry \cite{Watanabe2018,Hayami2018}. In addition, non-local order can arise from the itinerant property of electrons, for example $d$-wave Pomeranchuk instabilities in the presence of van Hove singularities (which is present in CeCoSi, see SM) \cite{Halboth2000_1, Yamase2007_1, Yamase2009_1}, even when only considering the Kramer's ground state. This instability can give rise to electric quadrupolar order, namely nematic order, or electric octupolar order if the order parameter is ferroically or antiferroically ordered, respectively \cite{Hitomi2016_1}. Thus, the spin and sublattice degrees of freedom combined with Fermi surface instabilities provide a rich environment for the formation of multipole orders, further enriched by hybridizing the quasi-quartet, and studying such quantum states may solve the mystery of the hidden order in CeCoSi.

With these multipole orderings in mind, we return to the open question of the nature of the HO state of CeCoSi. The interpretation of the NMR/NQR measurements relies on the formation of a quasiquartet between the ground- and first-excited states of the localized CEF scheme \cite{Yatsushiro2022_1, Manago2023_1}. In order to form this multipole order, an energy gap of 122\,K must be overcome to form the quasiquartet, which, versus $T_{HO}$=12\,K is a troubling difference in energy scales. A similar mechanism is proposed in CeRh$_2$As$_2$, however in this instance, the first excited state is much lower in energy at 30\,K, which is the same as the Kondo temperature and provides the way to form an effective quasiquartet \cite{Christovam2024_1}; this is not applicable in CeCoSi. Our CEF scheme vastly reduces the energy of the first excited state to less than 14\,K, which is compatible with the HO temperature. The scheme also provides a solution to the observed structural distortion related to the HO transition \cite{Matsumura2022_1}. By having a composite spatial state between the two cerium layers, our CEF scheme simultaneously provides the ingredients for multiple possible electronic quantum orders, and intrinsically provides a means to couple the electronic ordering and the structural instability, with the latter being caused by the former. Currently, in the context of our model, we cannot be more specific on the HO orderings, however, our model clearly provides new components with which to construct new quantum orders.

This phenomenon has been overlooked in $f$-electron systems, including the ones with locally broken centrosymmetry, such as CeRh$_2$As$_2$, UTe$_2$, UCoGe, URhGe, and UPt$_3$ \cite{Fischer2022_1}, and recently discovered Ce$_2$Ir$_3$Ga$_5$ \cite{Arushi2024_1}, which may be relevant when some particular structural arrangement and proper balance between ASOC and interlayer hopping are present. Naturally, in U-based compounds, for which spin-orbit coupling is stronger and the extension of the $5f$-orbital is larger, such effects might be at play while being more difficult to detect as compared to $4f$-electron compounds. Theoretical works in $f$-electron systems classify augmented or cluster multipolar possible states \cite{Nogaki2021_1, Suzuki2018_1, Hitomi2016_1, Hayami2018, Hayami2024_1} arising from this multisite physics. These predictions are for now based on a single pseudospin doublet CEF state and were lacking direct experimental confirmation. 
More broadly, as for multisite degrees of freedom, the entanglement of electron spin has proven effective to describe spin ladders \cite{Dagotto1992_1,Rice1993_1} or Shastry-Sutherland compounds \cite{Shastry1981_1}, but once again, this phenomenon appears much more scarce when considering the orbital degree of freedom; for example, few $d$-electrons systems forming orbital clusters or bilayers have been reported \cite{Singh2014_1, Rossi2021_1}. With the remarkably loose criteria for the emergence of hidden physics \cite{Zhang2014_1}, unexpected phenomena may be far more prevalent and varied than previously believed. In this paper, we have observed and evaluated the multisite CEF levels by Raman spectroscopy, paving the way to clarify quantum order with extended degrees of freedom. We hope that our work motivates new research into the entanglement of CEF and sublattice degrees of freedom in $f$-electron systems, and provides new avenues to pursue in this domain of hidden physics.


\clearpage 

%
\bibliography{BibFile} 
\bibliographystyle{sciencemag}

%
%
%
%
%
%


\section*{Acknowledgments}
We are grateful to Elena Hassinger, Gertrud Zwicknagl, Claudine Lacroix, Peter Thalmeier and Marie-Bernadette Lepetit for stimulating discussions. We would also like to thank Sylvia Kostmann for technical support.

\paragraph*{Funding:}
This work was supported by the European Research Council (ERC) under the European Union's Horizon 2020 research and innovation programme (Grant Agreement n$^{\circ}$ 865826). This work has received funding from the Agence Nationale de la Recherche under the project SEO-HiggS2 (ANR SEOHiggS2, Grant No. ANR-16-CE30-0014). This work has received funding from the German Science Foundation (DFG) through grant GE 602/4-1 Fermi-NESt. S. R. thanks the support from the Agence Nationale de la Recherche under the project SUPERNICKEL (Grant No. ANR-21-CE30-0041-04).

\paragraph*{Author contributions:}
M.-A. M., O. M., and C. G. conceived the project. O.M. and N. C.-C. grew the single crystals. O.M, and M.-A. M. performed the Raman measurements. G. G. performed the XRD measurements at the ESRF, while J. D. performed the Laue measurements. G.G. and S. R. analysed the X-ray data. C. F., A. P., O. M., and M.-A. M. performed the Raman measurements under high magnetic field. M.S. and Y.Y. performed the DFT calculations. O.M., Y.Y., and M.-A.M. interpreted the Raman results and co-wrote the paper with comments from all authors.

\paragraph*{Competing interests:}
The authors declare no competing interests. 

\paragraph*{Data and materials availability:}
All data are available in the main text or the supplementary materials.


\subsection*{Supplementary materials}
Materials and Methods\\
Supplementary Text\\
Figs. S1 to S7\\
Table S1 to S7\\
References \textit{(49-55)}\\ 


\newpage


\renewcommand{\thefigure}{S\arabic{figure}}
\renewcommand{\thetable}{S\arabic{table}}
\renewcommand{\theequation}{S\arabic{equation}}
\renewcommand{\thepage}{S\arabic{page}}
\setcounter{figure}{0}
\setcounter{table}{0}
\setcounter{equation}{0}
\setcounter{page}{1} 


\begin{center}
\section*{Supplementary Materials for\\ \scititle}

	Owen Moulding$^{1,2}$,
	Makoto Shimizu$^{3}$,
	Amit Pawbake$^{4}$, 
        Yingzheng Gao$^{1}$,\\
        Sitaram Ramakrishnan$^{1}$,
        Gaston Garbarino$^{5}$,
        Nubia Caroca-Canales$^{2}$,
        J\'er\^ome Debray$^{1}$,\\
        Clement Faugeras$^{4}$,
        Christoph Geibel$^{2}$,
        Youichi Yanase$^{3}$,
        Marie-Aude M\'{e}asson$^{1\ast}$\\
	\small$^{1}$Institut N\'{e}el CNRS/UGA UPR2940, 25 Rue des Martyrs, 38042 Grenoble, France.\\
	\small$^{2}$Max-Planck Institute for Chemical Physics of Solids, 01187 Dresden, Germany.\\
    \small$^{3}$Department of Physics, Graduate School of Science, Kyoto University, Kyoto 606-8502, Japan.\\
    \small$^{4}$LNCMI, UPR 3228, CNRS, EMFL, Universit\'{e} Grenoble Alpes, 38000 Grenoble, France.\\
    \small$^{5}$European Synchrotron Radiation Facility, 71 Avenue des Martyrs, 38000 Grenoble, France.\\
	\small$^\ast$Corresponding author. Email: marie-aude.measson@neel.cnrs.fr\\

\end{center}

\subsubsection*{This PDF file includes:}
Materials and Methods\\
Supplementary Text\\
Figures S1 to S7\\
Tables S1 to S7\\

\newpage


\subsection*{Materials and Methods}

Since CeCoSi forms in a peritectic reaction, we developed a cerium self-flux method for large single-crystal growth. In a first step, an appropriate amount of pure Ce, Co, and Si was melted in an arc furnace under an argon atmosphere to obtain a polycrystalline pellet. This pellet was ground, and the powder was inserted into a tantalum crucible and sealed. The crucible was then heated to 1250\,\si{\degree C} over 10\,hours before being slowly cooled to 800\,\si{\degree C} over 100\,hours. After that, the crucible was rapidly quenched by removing it from the furnace. This method provided millimetre-sized platelet single-crystals. 

We performed single-crystal x-ray diffraction at ambient conditions on the ID15B Beamline at the ESRF synchrotron in Grenoble, France. Data were obtained using omega scans from -90° to 90° with 0.5° step width, employing radiation with a wavelength of $\lambda$ = 0.4099\AA{} on the Eiger2X CdTe 9M PCD device detector. Intensities were integrated followed by scaling and empirical absorption correction with Laue symmetry 4/$mmm$ in CrysAlis \cite{Crysalis2023}. The resulting reflection file was imported into Jana 2006 for structure refinements \cite{Jana2006}.

Laue diffraction experiment has been implemented on a single-crystal with a laboratory tungsten source (set to 20\,kV and 30\,mA) equipped with a Photonic Science Laue camera. Patterns were recorded using an average function with a low acquisition time (20\,s). The sample was set into the diffractometer and confirmed to have large (00l) faces with perpendicular lateral edges. 

All Raman measurements were performed with a 532~nm solid-state laser with an incident laser power between 0.5-2\,\si{\milli\watt}. A Trivista 777 spectrometer equipped with ultra-low noise, cryogenically-cooled PyLon CCDs was used in both single-stage and triple-stage configurations. For the single-stage configuration, a RazorEdge filter cut the data below 70\,\si{\centi \meter^{-1}}, whereas in the triple-stage configuration, subtractive mode allowed measurements down to approximately 5\,\si{\centi \meter^{-1}}. High magnetic field data were acquired at the LNCMI, Grenoble, in a superconducting magnet with a maximum field of 14\,T. A single-stage Trivista 777 setup, and a filter cutting at approximately 50\,\si{\centi \meter^{-1}} measured the circularly polarised light.

Density functional theory calculations were fully relativistic and performed in a non-magnetic configuration using the full potential localised orbital (FPLO) basis~\cite{Koepernik1999}, and the generalised gradient approximation (GGA) exchange correlation functional \cite{Perdew1996}. We use the lattice constants and the atomic coordinates measured in reference \cite{Tanida2019_1} with a $12 \times 12 \times 12$ $k$-mesh. This non-correlated DFT based approach neglects the on-site U interaction acting on the $f$-states, resulting in the hybridisation between 4$f$ and other ligand states being overestimated, with the 4$f$-states' energy also being overestimated. However, the general tendencies, such as energy orderings, $j_z$ content, and van-Hove singularity in the electronic band dispersion, are well-captured.


\end{document}